\documentclass[10pt,conference]{IEEEtran}

\usepackage{cite}
\usepackage{amsmath,amssymb,amsfonts}
\usepackage{algorithmic}
\usepackage{graphicx}
\usepackage{textcomp}
\usepackage{xcolor}
\usepackage{comment}
\def\BibTeX{{\rm B\kern-.05em{\sc i\kern-.025em b}\kern-.08em
    T\kern-.1667em\lower.7ex\hbox{E}\kern-.125emX}}
 
\begin{document}

\title{Source Code Comments: Overlooked in the Realm of Code Clone Detection
}

\author{\IEEEauthorblockN{Sandeep Kaur Kuttal}
\IEEEauthorblockA{\textit{University of Tulsa} \\
sandeep-kuttal@utulsa.edu}
\and
\IEEEauthorblockN{Akash Ghosh}
\IEEEauthorblockA{\textit{University of Tulsa} \\
akashghosh@utulsa.edu}
}

\maketitle

\begin{abstract}
 Reusing code can produce duplicate or near-duplicate code clones in code repositories. Current code clone detection techniques, like Program Dependence Graphs, rely on code structure and their dependencies to detect clones. These techniques are expensive, using large amounts of processing power, time, and memory. In practice, programmers often utilize code comments to comprehend and reuse code, as comments carry important domain knowledge. But current code detection techniques ignore code comments, mainly due to the ambiguity of the English language. Recent advances in information retrieval techniques may have the potential to utilize code comments for clone detection. We investigated this by empirically comparing the accuracy of detecting clones with solely comments versus solely source code (without comments) on the JHotDraw package, which contains 315 classes and 27K lines of code. To detect clones at the file level, we used a topic modeling technique, Latent Dirichlet Allocation, to analyze code comments and GRAPLE -- utilizing Program Dependency Graph -- to analyze code. Our results show 94.86\% recall and 84.21\% precision  with Latent Dirichlet Allocation and 28.7\% recall and 55.39\% precision using GRAPLE. We found Latent Dirichlet Allocation generated false positives in cases where programs lacked quality comments. But this limitation can be addressed by using a hybrid approach: utilizing code comments at the file level to reduce the clone set and then using Program Dependency Graph-based techniques at the method level to detect precise clones. Our further analysis across Java and Python packages, Java Swing and PyGUI, found a recall of 74.86\% and a precision of 84.21\%. Our findings call for reexamining the assumptions regarding the use of code comments in current clone detection techniques.

\end{abstract}


\section{Introduction}
\label{sec:intro}

Code reuse involves modifying existing code fragments for a new context or problem and is a common practice among programmers to improve their productivity \cite{reuse1, reuse2, reuse3, reuse4, reuse5, reuse6, reuse7}. Although reuse practices by programmers can produce duplicate or near-duplicate fragments of code in code repositories \cite{Howard1994}. Studies have found 7\% to 23\% of software systems contain duplicated code fragments \cite{Baker1995, Baxter1998, Kapser2006, Mayrand1996}. Such duplicated code fragments -- code clones -- increase the complexity and cost of software maintenance \cite{Roy2014}. In many cases, code clones are also unintentionally generated by programmers as they work on similar programming tasks, overcome the limitations of programming languages, follow organizational coding conventions, or use design patterns \cite{Roy2007}. To support reuse and decrease maintenance costs, various code detection techniques have been utilized to find similarities between code fragments \cite{Roy2014, Bouktif, Adar, Bettenburg, Al-Omari2012, Bazrafshan, Chatterji, Cordy2003, Henderson }. 

One notable technique for detecting code clones is a Program Dependency Graph (PDG) \cite{Henderson}; especially to find semantic clones -- functionally similar code fragments that have same the pre and post conditions, but may or may not be syntactically similar. PDG based techniques are effective as they preserve statement ordering and analyze the data and control dependencies of the code \cite{Gabel}. But, these techniques are known to be computationally expensive and are biased, detecting certain clones but not others \cite{Henderson, Bellon, Roy2008, Sajnani2016, Wang2018, Saini2018}. 
 
 

Programmers often utilize code comments to understand and reuse code, as comments carry important domain knowledge \cite{Johnson, Seidl, Oezbek, Tenny, Woodfield}. But current code clone detection techniques only evaluate the source code and ignore comments within the source code, which are a significant portion of software systems \cite{Souza}. 

One of the primary reasons that clone detection techniques ignore code comments is the ambiguity of the English language. For humans, it is easy to comprehend the similarities and differences between words or topics, but a machine may treat the words differently. However, with recent advancements in topic modeling techniques, we are able to make more accurate predictions using machine learning and natural language processing tools. One of the most versatile topic modeling techniques is Latent Dirichlet Allocation (LDA), which has been extensively used for 
mining hidden topic patterns.
 

We conjectured that we can utilize code comments by applying current topic modeling techniques to detect semantic clones. Thus, we investigated the following:

\textbf{\textit{RQ1: Can code comments be utilized as effectively as source code to detect semantic clones at a file level?}}\\
To answer this, we empirically compared the precision and recall of LDA (solely using comments) and the PDG-based tool GRAPLE \cite{Henderson} (solely using source code) on an open source software package called JHotDraw (315 java source files with 27KLOC). We found 94.86\% recall and 84.21\% precision with LDA and 28.7\% recall and 55.39\% precision using GRAPLE.

\textbf{\textit{RQ2: Can we utilize a hybrid approach to detect clones?}} \\
Despite being able to use code comments to detect clones as in RQ1, our objective was not to replace current state-of-the-art semantic clone techniques such as PDG since they are more precise and generate fewer false positives. Hence, we formulated RQ2 to investigate if we can utilize the whole program file, both code comments and source code, to detect clones. Therefore, we analyzed the clone sets generated by LDA and PDG for the JHotDraw package and found common clones between them.

\textbf{\textit{RQ3: Can code comments help in detecting code clones across programming languages?}} \\
Since code comments are in plain English, they are more often similar across programming languages unlike the syntax of source code. Hence, we conjecture that good quality comments can support interoperability across several programming languages. To investigate the utilization of quality comments to detect clones in different languages, we formulated RQ3. Thus, we analyzed two popular Java and Python graphics packages, Java Swing (318 source files with 30KLOC) and PyGUI (355 source files with 28KLOC). When comparing across languages, we found recall of 74.86\% and precision of 84.21\%.

\section{Background}


\subsection{Program Dependency Graphs - Solely using Source Code}
Clone detection techniques utilize program representations such as source code text, tokens, abstract syntax trees (ASTs), and program dependence graphs (PDGs), each with their own set of advantages and disadvantages. For our purpose, PDG-based clone detection techniques are well suited and applied to detect semantic (functionally equivalent) clones as the use of PDGs preserves the semantics of statement ordering and are oblivious to code syntax.  

PDG is a static representation of the flow of data through a procedure\cite{Ferrante}. The nodes of a PDG could be declarations, simple statements, expressions, or control points of a source file. Control points are program branches, loops, or enter/exit points for a procedure. The edges of a PDG encode the data and control dependencies between program points. Since PDGs abstract many arbitrary syntactic decisions that a programmer makes while constructing a function, they are the best fit for finding semantic code clones \cite{Gabel}.

Program Dependence Graph G can be represented as a directed graph G with a set of vertices $V$ and a set of edges $E = V \times V$. A labeling function maps vertices or edges to labels $l$: V$\mid$ E $\rightarrow$ L. E can be represented by a matrix \textbf{E}. \textbf{$E_{i,j}$} = 1, if there is an edge between vertex $v_i$ to vertex $v_j$. Otherwise it is 0. 

A subgraph H of G (H subset of G) only exists if there is an injective mapping $m$: $V_H$ $\rightarrow$ $V_G$ such that:

\begin{enumerate}
\item Labels of all vertices in H should be the same when mapped to the labels of vertices in G
\item All edges in H are also in G 
\item Labels of all edges in H should map with all edge labels in G 
\end{enumerate}
These mappings $m$ are called embeddings. 
Fig.\ref{frag_pdg} shows two code fragments and their respective PDGs. Both PDG graphs are the same despite one code fragment using a for-loop and the other a while-loop to calculate the factorial of a number. 

\begin{figure*}
\centerline{\includegraphics[width = 10cm]{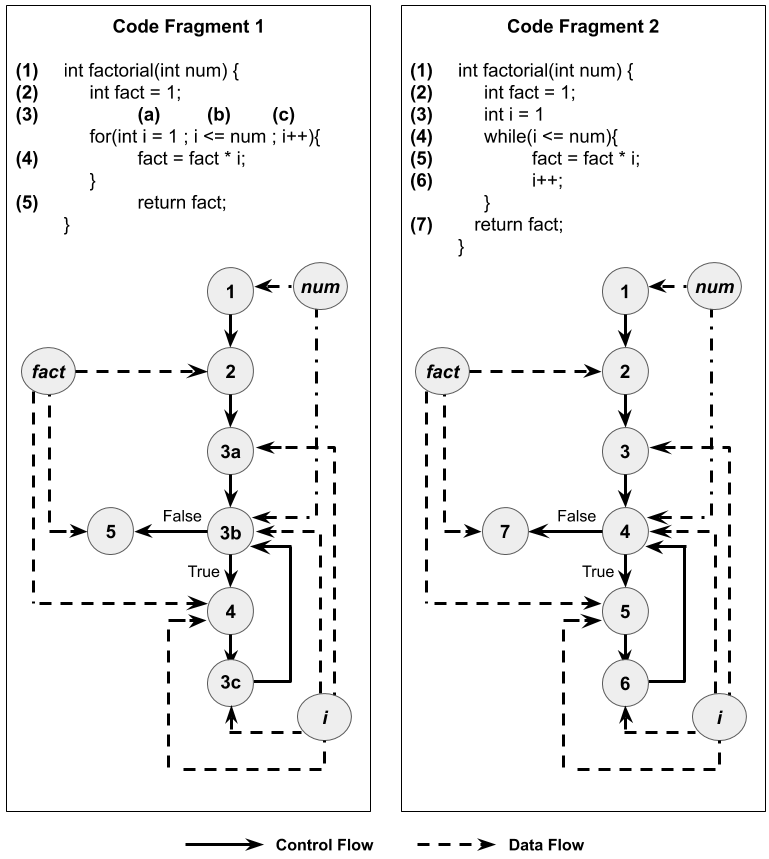}}
\caption{Code fragments for determining the factorial of a number, fragment 1 uses a for-loop and fragment 2 uses a while-loop. Each fragment is accompanied by a program dependency graph. Although the code fragments use different syntax, the dependency graphs are the same. Hence, the code fragments are semantic clones and can be detected by PDG-based techniques.
}
\label{code_frags}
\end{figure*}

\textbf{GRAPLE} (GRAph samPLE) is a code detection technique used to sample subgraphs of large graphs and then statistically estimate the subgraphs' characteristics \cite{Henderson}. We decided to use GRAPLE as it (1) is a statistically unbiased method for sampling dependence clones unlike most PDG-based clone detection tool that are biased towards detecting certain types of clones and (2) allows estimating parameters of the whole clone population to reduce computational cost as it is impractical to process all dependence clones \cite{Henderson}.


GRAPLE uses Frequent Subgraph Mining (FSGM). FSGM techniques identify recurring subgraphs of a graph database by traversing the frequently connected subgraph lattices that recur for $k > 1$, $k$ being the number of times it occurred in a subgraph. Once a subgraph is identified, its support is calculated as equal to the number of embeddings (mappings) that it has in the graph database. Finding reoccurring isomorphic subgraphs is an expensive technique, and it can be improved by storing the embeddings of each subgraph. Storing embeddings is called the canonicalization process. After the canonicalization process, GRAPLE randomly samples from the space of maximal frequent subgraphs. A frequent subgraph is maximal if no larger frequent subgraphs can be constructed. The sampling procedure allows us to compute selection probabilities for subgraphs, which can be used in statistical estimators such as the Horvitz-Thompson unequal probability estimator \cite{Thompson}. To use the Horvitz-Thompson probability estimator, it is necessary to determine the probability $p_i$ that the $i^{th}$ maximal frequent pattern $[[H_i]]$ is selected on a random walk of $k$-frequent connected subgraph lattice ($k$-$L_G$). GRAPLE uses Markov chains to compute the probability \cite{Grinstead}. A Markov chain moves from one state to another according to the probability $P_{i,j}$ (transition probability from state $s_i$ to $s_j$) in the transition matrix P of a finite set of states, S = {$s_1$, $\cdots$, $s_n$}. Here, states are considered as vertices of the lattice (i.e. frequent patterns $[[H_i]]$). The transition probability for an edge $v_i$ to $v_j$ is the reciprocal of the out-degree of $v_i$:

\[ P_{i,j}= \left\{ \begin{array}{lll}
 \frac{1}{\Sigma_k E_{i,j}} & &\mbox{if $E_{i,j}=1$}\\
          1 & \mbox{if $i=j $$\wedge$$ v_i $}&\mbox{is maximal}\\ 
        0&&\mbox{otherwise}.\end{array} \right. \]

\subsection{Latent Dirichlet Allocation - Solely using Comments}
Latent Dirichlet Allocation (LDA) is a statistical model for topic modeling that has been extensively used in natural language processing for representing text documents \cite{LDA}. To investigate the use of code comments in clone detection, we specifically utilized LDA as (1) it is the most popular topic modeling technique in the fields of machine learning and artificial intelligence \cite{wikipop, Hong2010}, (2) allows faster training, (3) is simple, and (4) efficiently utilize statistics. Further, we wanted to use a simple and efficient count based NLP technique to investigate our RQs rather than focusing on the performance as supported by sophisticated techniques like doc2vec. Given a corpus of documents, LDA identifies a set of topics; it associates a set of words with a topic, and a specific mixture of these topics for each document.


Basic terminologies to describe LDA are:

\textbf{Word:} A word is a basic unit that has been extracted from a vocabulary. 

\textbf{Document:}	A document is a series of words denoted by d = ${w_1, w_2, ..., w_n}$, where $w_n$ is the $n^{th}$ word in the series. We considered source files as documents. 

\textbf{Corpus:} A corpus is a set of M documents denoted by D = ${d_1, d_2, ....., d_M}$. We considered a code repository (or package) of source files as a corpus.

\textbf{Topic:} Topics are identified based on frequent and similar words in a corpus. Each document \textbf{d} can be modeled as a multinomial distribution $\theta(d)$ over T topics, and each topic $Z_j$, $j =1...T$ as a multinomial distribution $\psi(j)$ over the set of words W. 

Our task is to make an estimate of $\psi$ and $\theta$ in order to discover the set of topics used and the distribution of these topics in each document in a corpus of D documents \cite{LDA}. However, the LDA model assumes a prior Dirichlet distribution on $\theta$, thus allowing the estimation of $\psi$ without requiring the estimation of $\theta$.

The LDA algorithm \cite{LDA} works as follows:

    \begin{enumerate}
      \item Choose N $\sim$ Poisson($\xi$): Select the number of words N
      \item $\theta$$\sim${Dir($\alpha$)}: Select $\theta$ from the Dirichlet distribution parameterized by $\alpha$
     \item For each $w_n$ $\epsilon$ W do 
     \subitem - choose a topic $z_n$ $\sim$ Multinomial($theta$)
     \subitem - choose a word  $w_n$ from p($w_n$$\mid$$z_n$,$\beta$), a multinomial probability $\phi$$^{Z_n}$
    \end{enumerate}

\begin{figure*}
\centerline{\includegraphics[width = 15cm]{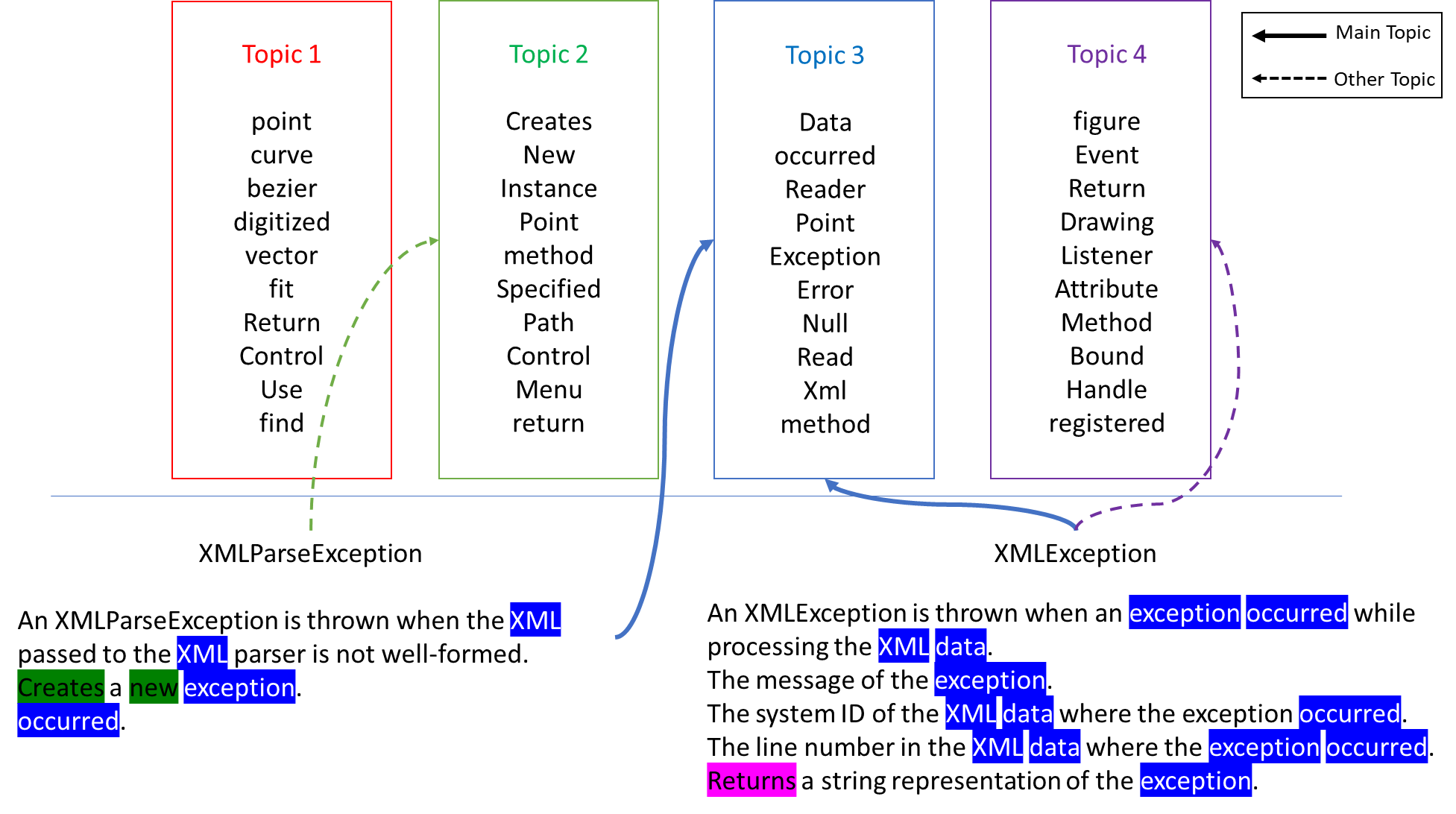}}
\caption{The mechanics of LDA model after the comments are extracted from source files. Bold arrows depict the highest relevance between the comments and a topic, whereas the dashed arrows depict the next best relevance between them. Both source files have the highest relevance with Topic 3; the highlighted words indicate the occurrence of each word in respect to their topic.}
\label{lda_model_motiv}
\end{figure*}

The LDA model uses a document word matrix $W_d$[w, $F_d$] = $n$, where $n$ is a value indicating the importance of the word $‘w’$ in the file $F_d$. The value of $n$ is computed using Gibbs sampling \cite{Griffiths}, which uses a Markov Chain Monte Carlo method to converge to the target distributions in some iterations. LDA assumes documents are produced from a mixture of topics, and these topics generate words based on their probability distribution. LDA samples topics of documents. LDA takes primarily two parameters, $\alpha$ and $\beta$, where $\alpha$ represents document-term density and $\beta$ represents topic-word density. The higher the value of $\alpha$, the more topics the documents are composed of. A lower value of $\alpha$ indicates that the document contains fewer topics. On the other hand, the higher the value of $\beta$, the more words the topic contains. A lower value of $\beta$ indicates that the topic contains fewer words. Fig. \ref{lda_model_motiv} shows the labelling of code comments from two source files and indicating the relevant topics.

\section{Approach}
\label{sec:approach}

To investigate the potential of using code comments to detect clones, we utilized Open Source Software (OSS) because code reuse is an acceptable practice in the OSS community as it is believed that knowledge should be shared with humankind \cite{Nakakoji}. 
 
\subsection{Dataset} 
We used three open source packages: JHotDraw \cite{JHD}, Java Swing \cite{JAVASwing}, and PyGUI \cite{PYGUI}. We used JHotDraw and Java Swing because they are popular packages that have been widely used in clone detection studies \cite{MCIdiff, Bouktif}. JHotDraw  contain 315 source files with approximately 27 KLOC and Java Swing contains 318 source files with 30 KLOC. To compare PDG and LDA, we used the JHotDraw package. For cross-language clone detection, we used Java Swing and PyGUI. We decided to use PyGUI as it is a graphical application similar to Java Swing making the two ideal for comparing cross-language clones. PyGUI contains 355 source files with 28 KLOC.

\subsection{System Configuration}
For computing and evaluating our dataset, we used a core i7 Quad Core 4th Gen processor with 3.6 GHz of clock speed and 32 GB of RAM.

\subsection{Procedure}
To study semantic clone detection, we conducted simulations using PDG and LDA. 


\subsubsection{Using PDG}
Program Dependence Graph is one of the most advanced procedures used for identifying semantic code clones. Our approach with PDG was accomplished with the help of GRAPLE, an existing code detection tool, which generates the clone sets by sampling isomorphic subgraphs. Fig. \ref{PDG} shows the overall mechanism for detecting clones using source code (without comments). The PDG evaluation consisted of:

\begin{figure}
\centerline{\includegraphics[width=0.5\textwidth]{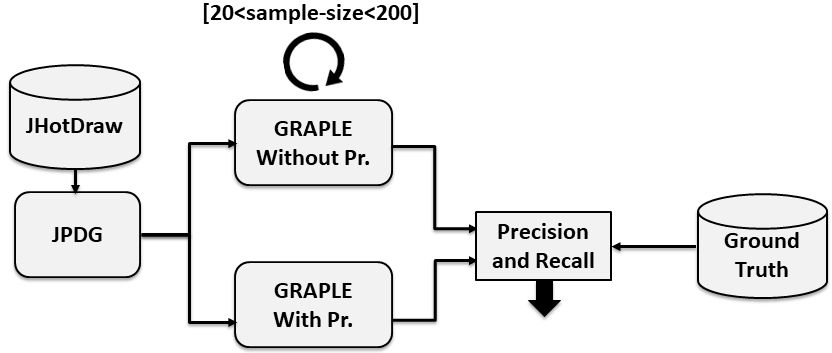}}
\caption{Block Diagram of PDG Evaluation.}
\label{PDG}
\end{figure}

\textbf{JPDG:} We used jpdg \cite{Henderson}, a PDG generator developed by Henderson et al. \cite{Podgurski} as it is more effective in detecting semantic clones over its predecessors.
The jpdg tool runs on Apache Buildr 1.4.15 and only works with Java 1.7 platform, so we had to rely solely on Java 1.7 for generating PDGs. jpdg generates a JSON file, which contains a list of dictionaries of two types: a list of vertices and a list of edges. The dictionaries for the vertices consist of key-value pairs, with keys such as``id,'' ``label,'' ``package\_name,'' ``class\_name,'' ``method\_name,'' ``type,'' ``start\_line,'' ``end\_line,'' etc. For instance, the key ``id'' represents the vertex number in the PDG. The dictionaries for the edges contained keys such as, ``src,''``targ,'' ``src\_label,'' ``targ\_label,'' etc. The jpdg generated the whole graph database in a single {\tt.veg} file (49 MB in size), which was used by GRAPLE for sampling out the clone sets with and without probability.



\textbf{GRAPLE with probability:} GRAPLE takes the {\tt.veg} file as an argument along with standard parameters such as minimum-support, sample-size, minimum-vertices, and probabilities. Then, GRAPLE creates a transition matrix for computing the selection probability $P_{i}$ of the frequent pattern $[[H_i]]$. Each cell of the transition matrix stores the probability of transitioning state from one vertex $v_{i}$ to another vertex $v_{j}$. Thus, this process encounters the ``Curse of Dimensionality,' as it consumes a huge amount of resources, both memory and processing power. To address this, we restricted to only generating subgraphs with $<$20 edges (still generating matrix of $2^{20}$ X $2^{20}$) as \cite{Henderson} indicated this to be when the submatrix is manageable. The selection probabilities were computed with support=5, sample-size=100, and min-vertices=8. Once selection probabilities were computed, GRAPLE generated the clone sets.


\textbf{GRAPLE without probability:} To avoid the ``Curse of Dimensionality,'' we used the option to turn off the selection probabilities. GRAPLE generated the {\tt maximal-patterns.dot} file containing digraphs that represented a clone set. We wrote a script in Python that identified all the clone sets from {\tt maximal-patterns.dot} by collecting the nodes of the digraph with labels and strings containing a source file. We varied the standard parameter, sample-size, from 20 to 200 and observed a very small increase in the clone sets.




The outputs from GRAPLE with and without probability were then parsed to the recall and precision module. The recall and precision for GRAPLE with selective probability was reported as $D_{reported\_GP}$ and without selective probability as $D_{reported\_GNP}$. 

\begin{figure}
\centerline{\includegraphics[width=0.5\textwidth]{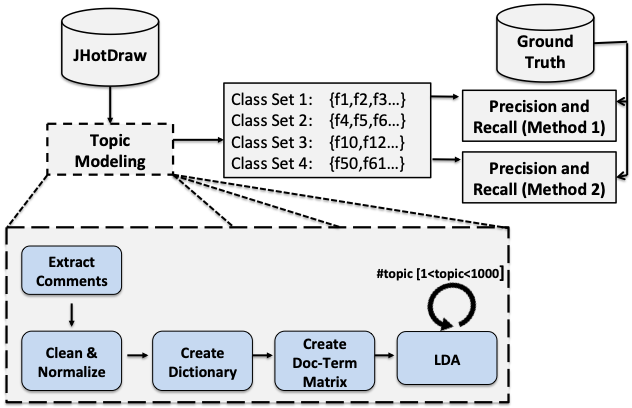}}
\caption{Block Diagram of LDA Evaluation.}
\label{LDA}
\end{figure}

\subsubsection{Using LDA}
 Fig. \ref{LDA} shows the block diagram of the evaluation mechanism, explaining the training of the LDA model and the extracting and generating of topics using it.

\textbf{Extracting Comments:} In order to train the model we needed to extract the comments from the source files. We wrote script in Python using RegEx to extract the comments from JHotDraw. 

\begin{figure}
\centerline{\includegraphics[width=0.5\textwidth]{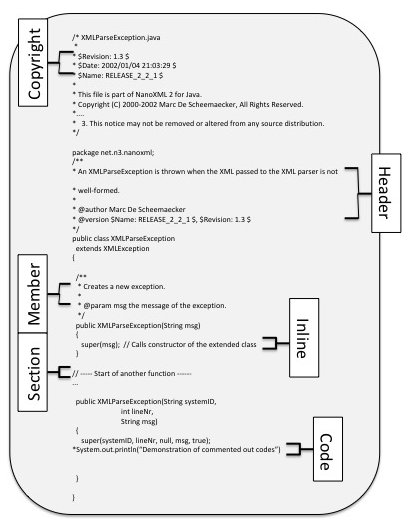}}
\caption{Different types of comments in a source file.}
\label{comment}
\end{figure}

Comments serve as an integral part of a source file, and are used for understanding code structure and functionality of the source file \cite{comment1, comment2, comment3}. Seidl et al. differentiated Java's code comments into seven categories \cite {Seidl}. We used all types of comments (refer Fig. \ref{comment}), except for copyright and task comments. The reason for this was that copyright comments do not contain information related to the functionality of the source code and task comments (developer notes containing todo) were not present in our dataset. 


Finally, we did not consider the HTML syntax from the comments. The primary reason for excluding the HTML tags (e.g. \textless html\textgreater,\textless p\textgreater, and \textless br\textgreater) was that they misdirected the LDA training process. As LDA uses multinomial distribution, the large frequency of the HTML tags in the corpus caused the model to assign higher probabilities to irrelevant characters.

\textbf{Clean and Normalize:} Once the comments were extracted from the source files, we used the Natural Language ToolKit to clean the stopwords and punctuations and then to normalize the comments. Once the comments were processed, they were combined together to form the corpus.

\textbf{Create Dictionary:} The corpus generated from the previous module was used to create a dictionary. A dictionary is a collection of all the unique words in the corpus. It also maps between the normalized words and assigned IDs (IDs are generated by the function itself). This was used later to train the LDA model. 

\textbf{Create Doc-Term Matrix:} Once a dictionary was prepared, it was used to create the document-term matrix. A typical document-term matrix displays the unique words in the columns and documents in rows. So, a cell $C_{i,j}$ in the doc-term matrix means the frequency of the $j^{th}$ word in the $i^{th}$ document. 

\textbf{LDA:} After the data was processed, we used the corpus  along with the dictionary to train the LDA model. Once the model was trained, individual source files were passed to it to generate an associated topic for that file. The model used passes and iterations between 1 to 1000 topics and then set to a specified value that generated the maximum number of topics. Our primary concern with LDA was that it cannot assign a meaningful label (topic) for the source files. Since we were more interested in the clusters of similar files assigned to a single topic (clone set), we were not concerned with LDA’s inability to assign meaningful names (topics). Below is an example of our output, with each clone set belonging to a topic.

{\tt Clone Set 1:[`Locator', `TriangleRotationHandler']}\newline
{\tt Clone Set 2:[`DrawApplet', `NetApplet', `SVGApplet', `PertApplet']}

We did not impose any restriction on the number of words for a labeled topic. 

We proposed two mechanisms to compare the clone sets
generated by LDA and the ground truth.

\begin{itemize}
	\item \textbf{Method 1}: Once the clone sets were generated, we calculated the precision and recall by unifying all the clone sets into one single clone set with unique file names. For instance, 
{\tt Clone Set 1:[`F1', `F2']}
{\tt Clone Set 2: [`F1',  `F4']} 
would become {\tt New Clone Set 1:[ `F1', `F2', 'F4']}. Similarly, with the ground truth we created another clone set and then performed the recall and precision as $D_{reported\_LDA1}$; similar approach by Maskeri et al. \cite{Girish}.
    \item \textbf{Method 2}: Here we kept the clone sets intact. We calculated the precision and recall between each individual clone set from $D_{reported\_LDA2}$ (clone sets reported by LDA) and $D_{actual\_LDA2}$ (clone sets reported by the ground truth), and then we took the highest value, since the pair with the highest value would undoubtedly be a match. Finally, we took the average of precision and recall for all the clone sets of $D_{reported\_LDA2}$.
\end{itemize}


\subsection{Ground Truth}
				
To evaluate the effectiveness of the PDG and LDA methods, a senior undergraduate and graduate student investigated the semantic clones using Java Compare and Eclipse Java Editor to build the ground truth. In total, 45 hours were spent generating the ground truth $D_{actual}$. For cross-language, another senior undergraduate and graduate student created the ground truth. They spent 55 hours generating the ground truth. 

In the JHotDraw package, 52 clone sets were found; similarly, in Java Swing, 19 clone sets were found, and in PyGUI, 50 clone sets were found. The clone sets consisted of clones ranging from a minimum of 2 to a maximum of 45.

\textit{Precision} is the percentage of correctly reported differential multisets. Precision is calculated as 	
$\mid$$D_{reported}$ $\cap$ $D_{actual}$$\mid$ $/$ $\mid $$D_{reported}$$\mid$, where $D_{reported}$ is the set of multisets reported by either LDA or PDG.  

\textit{Recall} is the percentage of actual differential multisets reported. It is calculated as $\mid$$D_{reported}$ $\cap$ $D_{actual}$$\mid$ $/$ $\mid$$D_{actual}$$\mid$.

\section{Results}
\label{sec:results}

The state-of-the-art semantic clone detection techniques  \cite{LICCA2018, clcminer} rely on programs that yield the same outputs using dynamic code similarity detection \cite{Lxiang, shihan, kim}, and identify similar behaviors of different programs by comparing instruction-level execution \cite{fhsu}. These approaches are precise, but not scalable, and have limitations for practical usage. Moreover, none of these approaches have considered code comments for identifying similar code fragments. We wanted to explore the feasibility of using code comments as a parameter for clone detection at a file level. Hence, we investigated:\\

\noindent\textbf{\textit{RQ1: Can code comments be utilized as effectively as source code for detecting semantic clones at a file level?}}


 \subsection{GRAPLE}
 To investigate RQ1, first, we used GRAPLE \cite{Henderson} with and without the selection probability $P_{i}$ (refer Sec.~\ref{sec:approach}).

\textbf{GRAPLE with Probability:} We used jpdg to generate the program dependence graph and then used GRAPLE to detect the clone sets. jpdg generated the {\tt.veg} file for JHotDraw, which consisted of 61K vertices and over 118K edges, an overall size of 49 MB. We used the same parameters as used by Henderson et al. \cite{Henderson}: with the sample-size set to 100, the minimum vertices set to 8, and the support set to 5. Using the selection probabilities generated 80 clone sets. 
By manually inspecting all clone sets, we found most had just one source file. These files were removed from our $D_{reported\_GP}$; as GRAPLE detected similar method/function-level clones within that same file. These files were also excluded from the clone sets, as we were interested in detecting clones at the file level. Moreover, we also found some clone sets that were duplicates, where the file pairs had more than one similar functionality on different sections of their code. These extra clone sets were removed and only one was retained as it did not matter if more than one part of a source file was similar to another part of the source file.  After excluding the function-level and duplicated clone sets, only 22 remained. We found $D_{reported\_GP}$ recall of 28.7\% and precision of 55.39\%.

\textbf{GRAPLE without Probability:} Without using the selection probabilities, we generated clone sets by varying the sample-size from 20 to 200, however, only a small increase in clone sets were observed. Hence, we generated clone sets using the above mentioned specifications: sample-size set to 100, minimum-vertices to 8, and support set to 5. We found a total of 68 clone sets were generated. After excluding the function-level and duplicated clone sets, only 17 remained. We found $D_{reported\_GNP}$ recall of 27.84\% and precision of 52.94\%. 

\subsection{LDA}
Secondly, to study whether comments can be utilized for clone detection, we trained the LDA model on code comments from 310 files of JHotDraw using Method 1 and Method 2. Five files were excluded from the data set as they did not contain any code comments.

\begin{figure}
\centerline{\includegraphics[width=0.5\textwidth]{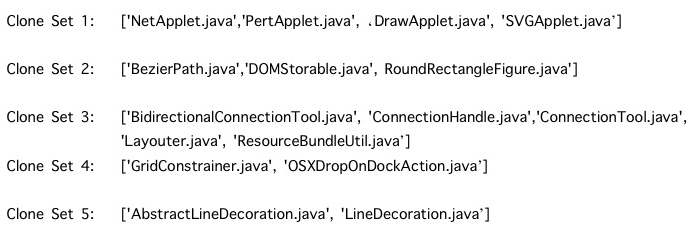}}
\caption{Snapshot of Clone Sets of JHotDraw Package.}
\label{cloneset}
\end{figure}

\textbf{LDA using Method 1:}
We used LDA to generate topics (interpreted as a set of semantically related linguistic terms) derived from comments. The LDA model was trained using all the files in the corpus. We randomly set the topic limit to 100. Using this process, 66 clone sets containing 274 files were extracted. Fig. \ref{cloneset} depicts a subset of clone sets generated by LDA (Note: since the assigned topics were not labeled by LDA, we refer to them as clone sets). Once all the clones sets were generated, precision and recall metrics were computed. Based on the $D_{reported\_LDA1}$, recall was 94.86\% and precision was 84.21\%. Hence, it was concluded that we can utilize comments to detect code clones.

\textbf{LDA using Method 2:}
 In Method 2, we used a more sophisticated approach to calculate the precision and recall. The topic number was set to 105 as Ghosh and Kuttal \cite{akash2018} found  when LDA was trained with over 1-1000 topics using 1000 iterations with 50 passes, it had highest recall at 100-108 topics. LDA found 7 clone sets with 21 files. Based on the $D_{reported\_LDA2}$, recall was 28.61\% and precision was 88.57\%. After manually analyzing the codes to check the authenticity of the clone sets, we found that the matched clone sets, i.e $D_{reported\_LDA2}$ $\cap$ $D_{actual\_LDA}$, were functionally identical in terms of object or instance creation.


\begin{table}[h]
    \caption{Recall and Precision for JHotDraw}
    \begin{center}
        \begin{tabular}{|c|c|c|c|c|}
        \hline
        &&\textbf{\#Clones}&\textbf{Recall}&\textbf{Precision} \\
        \cline{1-5} 
        \cline{1-5} 
        \hline
          \textbf{} & \textbf{\textit{With Pr.}}&22&28.7\%& 55.39\% \\
            
        \cline{2-5} 
          \textbf{PDG} & \textbf{\textit{Without Pr.}}&17&27.84\%&52.94\%  \\
        \hline
        \hline
            \textbf{} & \textbf{\textit{Method 1}}&66&94.86\%&84.21\% \\
        \cline{2-5} 
            \textbf{LDA} & \textbf{\textit{Method 2}}&7&28.61\%&88.57\% \\
        \hline
        
        \end{tabular}
    \label{jhotdraw}
    \end{center}
\end{table}

\subsection{Discussions}
Table \ref{jhotdraw} summarizes the number of clone sets reported, the recall, and the precision of PDG (with and without probability) and LDA (with the different methods). Our results show that we can utilize comments for detecting code clones at a file level. 

\textbf{With vs. Without the Selection Probabilities:} We wanted to compare and contrast the clone sets generated with and without the selection probabilities. Table \ref{jhotdraw} shows that 22 clone sets were detected using selection probabilities and 17 without using selection probabilities. To check the quality of the clones, we manually went through all the clone sets reported between the two approaches and also measured the correlation between $D_{reported\_GP}$ by GRAPLE with selection probabilities and $D_{reported\_GNP}$ by GRAPLE without selection probabilities. It was found that 16 out of 17 clone sets reported by $D_{reported\_GNP}$ had been also reported by $D_{reported\_GP}$. In addition, we found that without selection probability GRAPLE missed 5 clone sets and wrongly classified 1 clone set compared to including the selection probabilities. Although the $D_{reported\_GP}$ produced a comprehensive clone set, it cost 30 hours and 74 GB of memory. On the other hand, $D_{reported\_GNP}$ found 17 clone sets in 4.5 seconds and consumed 481.5 MB.   

\textbf{Effect of PDG based Tool's Constraints:} The PDG's (GRAPLE) performance was attributed to the parameters, code structure, and code dependencies. For example, PDG was not able to detect the clone set {\tt[XMLParseException, XMLException]} as clones because while comparing these files, PDG considered the difference in the number of arguments, classes extended (different super classes), and the presence of additional methods.  {\tt XMLParseException} extends {\tt Runtime Exception} and requires three argument values: {\tt`name,' `message,'} and {\tt`LineNr.'}  The {\tt XMLException} class extends {\tt Exception} and requires five arguments in the constructor: {\tt`SystemID,' `lineNr,' `Exception,' `msg' }and {\tt`reportParams'} and has a separate method to print stack traces for the exceptions.  These constraints by GRAPLE affected the detection.


\textbf{Recall vs. Precision using LDA:} Balancing the trade-off between recall and precision is one of the major concerns in using LDA. When we applied the LDA model to the comments and varied the topic numbers from 1 to 1000, we observed a steady increase in precision and decrease in recall. Identifying the range where the recall and precision will be balanced is challenging and may differ based on package size and contents of the comments. 

\textbf{Effect of Code Comments on LDA:} The performance of LDA was very much dependent on the quality of the code comments. When the code comments were present LDA was able to detect clones that were not detected by PDG. For example, LDA found the {\tt[XMLParseException, XMLException]} clone set, which PDG did not find. As seen in Fig. \ref{lda_model_motiv}, the code comments of both source files contained better information on the exceptions. Although, in the cases where the code comments were vague, LDA detected false positives. Moreover, as expected, the lack of code comments resulted in LDA missing clones completely.

The problem of mismatched, missing, and outdated comments has been well approached by the software research community \cite{McBurney2014,Sridhara2010} by utilizing techniques such as manually crafted heuristics and stereotypes \cite{Moreno2013}, information retrieval \cite{Haiduc2010a, Haiduc2010b},  probabilistic models \cite{Gu2016, Loyola2017,Wang2016, White2016}, Recurrent Neural Networks \cite{Iyer2016}, and deep learning \cite{hu2018}, we believe that these advanced code generation techniques can be utilized to address LDA's limitation of vague comments or lack of comments. \\

\noindent\textbf{\textit{RQ2: Can we utilize a hybrid approach to detect clones?}}\\

To investigate whether we could utilize the whole source file, i.e. both the source code and its comments, we started by examining the similarities between the clone sets generated by LDA and PDG for the JHotDraw package. We computed $\mid$$D_{reported\_LDA1}$ $\cap$ $D_{reported\_GNP}$$\mid$, where $D_{reported\_LDA1}$ is clones reported by LDA using Method 1 and $D_{reported\_GNP}$ is clones reported by GRAPLE without selection probability. We started by setting the similarity index to 1 ($S_1$), i.e. all the clone sets with at least one matching file. For instance, {\tt`BezierPath'} is the matching file among $D_{reported\_LDA1}$ = {\tt[`BezierPath,' `DOMStorable,' `RoundRectangleFigure']} and $D_{reported\_GNP}$ = {\tt[`BezierPath,' `DoubleStroke']}. Next, we set the similarity index to 2 ($S_2$) and 3 ($S_3$). By following this procedure, we found that the largest clone set can be generated when the similarity index is 3, making it the maximum similarity index. 

 After collecting clone sets based on the similarity indices, we created a superset containing all those files ($S_1 \cap S_2 \cap S_3$). The superset contained unique file names. In our case we found 40-50 common files within the LDA and PDG clone set. This superset can be used as input for PDG based techniques like GRAPLE to detect function level clones. This will help in utilizing code as well as the comments of a program to find unique sets of clones.

\subsection{Discussions}
 We recommend the use of a hybrid technique, i.e. utilizing both LDA and PDG to detect the semantic clones. By applying LDA, we obtained 130 unique files for our data set\footnote{LDA generates the clone sets based on the random states assigned. For our experiment, we set the random seed to 100, so that the LDA model always returned the same number of clone sets; i.e., the same number of unique files.}, which reduced the dataset size by more than 50\%. This reduced dataset obtained from LDA can be used by GRAPLE with selection probabilities. Hence, the hybrid approach can reduce both the time and space complexity of the whole process. This will help in utilizing source code as well as the comments of a program to detect unique sets of clones. Processing all clone dependencies for even moderately sized programs is impractical. As noted by Henderson et al.\cite{Henderson}, for programs with 70 KLOC, around 10 million clones were detected before the space was exhausted. LDA can be utilized to generate the clone sets at the file level and then PDG-based techniques can be applied on these selected clone sets to detect the function-level clones. \\

\noindent\textbf{\textit{RQ3: Can code comments help in detecting clones across programming languages?}}\\


To determine whether comments can be utilized to detect clones across multiple programming languages, we investigated code comments in Java Swing and PyGUI which are popular options in Java and Python, respectively, to build graphical user interfaces. We used Method 1 and Method 2 as discussed in Section \ref{sec:approach}. Table \ref{crosslanguage} summarizes the recall and precision of Java Swing and PyGUI.  For Java Swing, we found 66 clone sets with a 90.68\% recall and 49.49\% precision according to Method 1 and 69 clone sets with a 37.83\% recall and 32.77\% precision according to Method 2. For PyGUI, we found 61 clone sets with 51.12\% recall and 39.21\% precision for Method 1 and 58 clone sets with 65.62\% recall and 53.33\% precision for Method 2.

In order to detect the clone sets across the two packages Java Swing and PyGUI using LDA, we created a dictionary using both of the packages. The corpus consisted of 314 source files from Java Swing and 355 source files from PyGUI. For each source file, comments were extracted out and parsed to the LDA model one at a time such that each source file was assigned to a particular topic number. Source files with similar topic numbers were put together to form a clone set. Once the clone sets were created, we calculated the recall and precision. We found 88 clones common between Java Swing and PyGUI (refer Table \ref{crosslanguage}), with recall of 74.86\% and precision of 84.21\% using Method 1 and  we found recall of 28.61\% and precision of 58.57\% using Method 2.

\subsection{Discussions}
 Our results show that we can utilize code comments to detect clones across different languages to an extent. But limitations of the quality of code comments and large number of false positives will persist. As discussed before, we recommend using automated code comment generation and hybrid-technique utilizing LDA and language specific PDG to detect clones.

\begin{table}[htbp]
\caption{Cross-Language Evaluation of Java Swing and PyGUI}
\begin{center}
\begin{tabular}{|c|c|c|c|c|}

\hline
&&\textbf{\#Clones}&\textbf{Recall}&\textbf{Precision} \\
\cline{1-5} 
\cline{1-5} 
\textbf{} & \textbf{\textit{Method 1}} &66 &90.68\%  &49.49\% \\
\cline{2-5} 
\textbf{Java Swing} & \textbf{\textit{Method 2}}  &69 &37.83\%  &32.77\% \\
\hline
\textbf{} & \textbf{\textit{Method 1}} &61 &51.12\%  &39.21\% \\
\cline{2-5}
\textbf{PyGUI} & \textbf{\textit{Method 2}}  &58 &65.62\%  &53.33\% \\
\hline
\hline
\textbf{} & \textbf{\textit{Method 1}} &88 &74.86\%  &84.21\% \\
\cline{2-5} 
\textbf{Java Swing - PyGUI} & \textbf{\textit{Method 2}}  &81 &28.61\%  &58.57\% \\
\hline
\end{tabular}
\label{crosslanguage}
\end{center}
\end{table}

\section{Threats to validity}

\textbf{Threat to External Validity:} 
We studied medium-sized Java projects and a Python project, which cannot serve as an exemplar for all software systems. Secondly, for cross-language verification we studied only two languages, Java and Python. Thirdly, we explored only one PDG based tool -- GRAPLE -- and only one topic modeling technique -- LDA. Despite these limitations, this study is a first step towards exploring the viability of the utilization of code comments in clone detection. Future studies on large-scale systems and with different languages, tools, and techniques need to be done to analyze the generalizability of our results.

\textbf{Threat to Internal Validity:} 
LDA's recall and precision depends largely on the quantity and quality of contents. In our data, we adjusted the number of passes and iterations to find a balance between the recall and precision of LDA. But, in practice, finding the right balance is challenging and limits the usage of LDA. Additionally, the LDA approach cannot be applied to visual programming languages as they do not contain code comments. Yet, our results indicate that topic modeling techniques could be applied to text based programming languages to detect clones. 




\textbf{Threat to Construct Validity:} 
To maintain consistency throughout the study, we analyzed only a single version of the Java library files. This decision was based on the facts that GRAPLE and past studies used JHOtDraw v7.0.6. Furthermore, with different versions, the number of files could have been altered by the authors, and thus might have caused mis-matches in the detection process. 

\section{Related Work}
\label{sec:relatedwork}


\subsection{Code Clone Detection Techniques}
In software engineering, many techniques \cite{Roy2014} have been proposed to detect code clones based on token similarity (e.g., CCFinder \cite{Bouktif}, CloneMiner \cite{Adar}, and CloneDetective \cite{Bettenburg}), Abstract Syntax Tree (AST) similarity (e.g., CloneDR \cite{Al-Omari2012}, Deckard \cite{Bazrafshan}), or Program Dependence Graph similarity (e.g., \cite{Chatterji, Cordy2003, Henderson}). These clone detectors can detect not only textually identical clones (Type I), but also parameterized clones (Type II) and gapped clones (Type III) \cite{Asaduzzaman2012}. Textually identical clones refer to code fragments with differences only in whitespace, layout, and comments. Parameterized clones refer to syntactically identical code fragments, except for differences in identifiers, literals, and types. Gapped clones refer to copied fragments with further modifications such as changed, added, or removed statements. A code clone often appears in multiple places in the system; i.e., it has multiple instances. Detecting and analyzing differences in parameterized and gapped clones has been used in the software engineering literature to manage and maintain code clones by identifying refactoring opportunities \cite{Gode}, detecting bugs \cite{Bellon}, supporting change propagation in code clones \cite{Baxter, Cordy2011}, searching code \cite{Sirres}, and detecting plagiarism \cite{Facoy2018}.  

Despite this detection and analysis, finding Type III along with semantic clones (Type IV) is still an open research problem \cite{Roy2014}. Semantic clones are functionally similar code fragments that have similar pre and post conditions, but may or may not be syntactically similar. Basit et al. \cite{Basit} have explored the applicability of generics for the removal of code clones in the Java Buffer Library and the C++ Standard Template Library (STL) and concluded that programming language constructs limit the applicability of generics or templates for clone removal. Most existing clone detection techniques analyze the lower level code (e.g., assembly code, Java Bytecode, or .Net intermediate language) as obtained from the transformation by the compiler rather than from analyzing the textual source code\cite{Al-Omari,Keivanloo, Davis}.


Clone detection mechanisms have utilized various techniques, like searching for isomorphic sub-graphs \cite{Komondoor}, tracing program executions \cite{Gabel}, using deep learning \cite{white}, and using abstract memory states \cite{Facoy2018}. Overall, prior research has ignored code comments, therefore, we explored the viability of comments to detect semantic clones at a file level by using LDA, a topic modeling technique.

\subsubsection{Clone Detection using PDG}
Krinke’s Duplix algorithm \cite{Krinke} and Komondoor and Horwitz’s algorithms \cite{Komondoor} had been utilized as PDG-based techniques to detect clones. Higo and Kusomoto extended Komondoor's algorithm to detect contiguous clones \cite{Higo2009, Higo2011}. Deckard \cite{Bazrafshan} showed an innovative way to map PDGs to abstract syntax trees for detecting clones. Pham et al. \cite{Pham} conducted research to detect clones by using labeled directed graphs and finding clones with vSiGram. Henderson and Podgurski \cite{Henderson} developed a PDG-based clone detection tool using maximal frequent subgraph mining with different graph mining patterns \cite{Cheng} and GRAPLE \cite{Henderson}. We used the GRAPLE semantic clone detection tool to detect clones when considering code without comments.

\subsection{Clone Detection Across Languages}
The problem of detecting clones persists across multiple languages especially in large-scale software systems. With an increase in the size of clones the relation between them gets more subtle \cite{Sdang, survey, duplicate, quality}. The existing approaches as mentioned above perform clone detection only in a single language. Some research has been conducted on cross-language clone detection. Kraft et al. \cite{kraft} conducted clone detection research mainly on .NET languages. Microsoft's Common Intermediate Language (CIL) has been used by Al-Omari et al. \cite{Al-Omari2012} to represent source code, which detects similar code fragments. This tool is restricted to find true positive cross-language code clones in .NET languages only. Another important contribution comes from Avetisyan et al. \cite{avetisyan}, which uses LLVM bitcode to detect semantic clones. The approach is applicable to any languages that can be compiled to LLVM bitcode. Cheng et al. \cite{clcminer} conceptualized the notion of detecting similarities in sets of components written in different languages, using Natural Language Processing techniques to mine projects' revision histories. Vislavski et al. \cite{LICCA2018} designed a tool, LICCA, for cross-language clone detection that is based on intermediate program representation to unify semantically similar code fragments. They evaluated the tool on an extended set of cloning scenarios over five different languages (i.e. Java, JavaScript, C, Modula-2, Scheme). One of the limitations of LICCA \cite{LICCA2018} is that it can detect small fragments (few LOCs) of code clones and is ineffective in large software systems. None of these techniques consider code comments for clone detection.

\subsection{Software Engineering and LDA}
LDA has been widely used in software engineering but mostly for program comprehension and maintainability. Wilde et al. \cite{Wilde} proposed the use of linguistic information to identify the functional intent of the system. Biggerstaff et al. \cite{Biggerstaff} have suggested the assignment of domain concepts as an approach to program comprehension. Prior research has proposed using function names and signatures to obtain domain specific functions \cite{Caprile}. Furthermore, file names often carry the functional intent of the source code specified in the file \cite{Anquetil}. Antoniol et al. used information retrieval methods to find traceability links between code documentation and source code  \cite{Antoniol 2012}. Oezbek et. al\cite{Oezbek} created an Eclipse plugin, JTourBus, to lead the programmer directly to relevant details by creating a tour through the source code.
Kuhn et al. \cite{Kuhn} used a Latent Semantic Analysis based approach for software comprehension identifying topics in source code by semantically clustering software artifacts such as methods, files, or packages based on identifiers' names and  comments. 
Unlike these approaches, we are interested in utilizing linguistic topics (of code comments) to detect clones rather than comprehend programs (source code).

\section{Conclusions}
This is the first study to investigate the realm of code comments to detect code clones. We made the following contributions:
\begin{enumerate}
  \item Provided empirical evidence that code comments can be utilized for detecting clones at a file level and even across programming languages. We found that the precision of the detected clones largely depended on the quality of the comments. In the presence of vague or incomplete comments, there was a higher number of false positives. 
 
  \item Demonstrated that clone detection may utilize a hybrid approach, a combination of LDA and PDG, first detecting clones at the file level using code comments and then at the method or statement level using the source code. A hybrid approach can help in reducing the cost of clone detection by utilizing less resource-intensive techniques, like LDA which can be applied to reduce the number of clone sets.  It also helps in increasing accuracy of clone detection by utilizing sophisticated and resource-intensive techniques, like PDG which can utilize code structure for finding method and statement level clones. 
  
  \item Revealed that PDG-based techniques can miss detecting some clones because of their strict constraints, like matching parameters, code structure, and code dependencies.
 
  \end{enumerate}
 Our study provided evidence that comments, which are underrated in clone detection research, can be effectively utilized.

\newpage{\pagestyle{empty}\cleardoublepage}

\end{document}